# Exploiting Dynamically Propositional Logic Structures in SAT

Jingchao Chen

School of Informatics, Donghua University
2999 North Renmin Road, Songjiang District, Shanghai 201620, P. R. China
chen-jc@dhu.edu.cn

**Abstract.** The 32-bit hwb (hwb-n32 for short) problem is from equivalence checking that arises in combining two circuits computing the hidden weighted bit function. Since 2002, it remains still unsolvable in every SAT competition. This paper focuses on solving problems such as hwb-n32. Generally speaking, modern solvers can detect only XOR, AND, OR and ITE gates. Other non-clausal formulas (propositional logic structures) cannot be detected. To solve the hwb-n32 problem, we extract dynamically some special propositional logic structures, and then use a variant of DPLL-based solvers to solve the subproblem simplified by the extracted structure information. Using the dynamic extraction technique, we solved efficiently the hwb-n32 problem, even some of which were solved within 3000 seconds.

## 1 Introduction

In practice, although numerous SAT instances have already been solved, there are yet many unsolvable SAT instances. For example, the hwb-n32 family is an example that cannot be solved in every SAT competition. This paper focuses on solving problems such as hwb-n32. A SAT problem is generally represented in Conjunctive Normal Form (CNF), in which a formula consists of a conjunction of clauses, each clause being a disjunction of Boolean literals, where each literal is a positive or negated Boolean variable. The structure of the hwb-n32 problem appears to be simple, since it consists of a small amount of variables and CNF clauses. However, what is its structured information is still a puzzle.

Exploiting structured information can be traced back to the work of Warners et al. in 1998 [4]. They made use of a linear programming technique to detect XOR constraints, and solved successfully the DIMACS 32-bit parity problem. Since then, many other exploiting approaches have been proposed. In 2000, Li [5] made use of lists of equivalent variables to detect XOR constraints whose length is smaller or equal to 3. His approach is time-consuming. To avoid useless syntactical tests, Ostrowski et al. [1] suggested making use of an original concept of partial graph. Recently, Chen [6] presented a simpler approach, which collects together clauses with the same variable set, and then detect XOR constraints by counting the number of negated variables in each clause, since XOR constraints follow an odd or even rule with respect to the number of negated variables. These approaches above are restricted to extract only XOR constraints, except that the approach of Ostrowski et al. can extract other propositional logic such as AND, OR gates. Another common feature of these approaches is that they all exploit statically structured information. That is, they graft structured information exploiting as a preprocessing step to the DPLL procedure. The PrecoSAT solver [9], the Gold Medal winners in the application category of the SAT 2009 competition, can detect dynamically XOR, AND, OR and ITE gates, but does not support the conflict analysis of these gates during the search. The extracted gates are used to simplify the CNF formula and remove the redundant variables.

The goal of this paper is to exploit other structured information that cannot be expressed as XOR, AND, OR and ITE gates. In theory, identifying dependence relations between variables from a CNF formula is time-consuming in the general case, since this has been shown to be coNP-complete [7]. However, because our exploiting is restricted to a special case, neither a general case nor dependence relation, and the test is syntactical, not semantic, the cost is low. We found that a special propositional logic is useful to speed up the solving procedure. However, the special propositional logic is not contained in initial CNF inputs, but occurs during the search process. Therefore, to make use of the special propositional logic to speed up the solver, we have to extract dynamically structured information. The new dynamical exploiting technique is helpful to understand the nature of some special SAT instances. Based on our experimental results, using the dynamical exploiting technique, we can solve hwb-n32, which cannot be solved so far by any solver.

## 2 Reasoning Rules with Propositional Logic extracted from a CNF formula

Some propositional logic structures are often hidden in some SAT problems. How to detect general propositional logic structures from a CNF formula is a challenge, even if it originates from a propositional logic structure. In this paper, we focus on detecting some special logic structures. Generally speaking, solving a CNF is more efficient than solving propositional logic. However, we found that handling directly propositional logic is sometimes more efficient. The example is reasoning rules extracted from a CNF encoding of equivalence checking problem of two different circuits computing the hidden weighted bit function [10]. Let $HWB_n : \{0,1\}^n \to \{0,1\}$ denote the Hidden Weighted Bit function. In details, $HWB_n$ is defined as $HWB_n(x_1, x_2,\ldots,x_n) = x_{sum}$, where $sum = x_1 +\ldots+ x_n$ and $x_0 = 0$. That is, the function selects $x_i$ as its output if $sum = i$, i.e. $i$ inputs are 1. Here are reasoning rules extracted from a $HWB$ CNF encoding, where $\alpha$ is a sequence of clauses (including XOR constraints), $\beta$ is propositional logic, and $\alpha \vdash \beta$ denotes that $\beta$ is a logical consequence of $\alpha$. The rules here will be used in our solver. Notice, in some premises, we use directly XOR constraints, not CNF clauses. This is because XOR constraints are easily detected from a CNF formula.

| Clauses | | logical consequence |
|---|---|---|
| (1) $(A \oplus B \oplus C = 1)$<br>$(A \vee D)(\neg B \vee D)$ | $\vdash$ | $C$ imply $D$ |
| (2) $(A \oplus B \oplus C = 1)$<br>$(A \vee B)(A \vee C)$ | $\vdash$ | $A$ |
| (3) $A \oplus B \oplus C = 1$<br>$(A \vee B)(A \vee C)(B \vee C)$ | $\vdash$ | $A \wedge B \wedge C$ |
| (4) $A \oplus B \oplus C \oplus D = 1$<br>$(A \vee B)(A \vee C)(B \vee C)$ | $\vdash$ | $\neg D$ imply $A \wedge B \wedge C$ |
| (5) $(A \oplus B \oplus C = 1)$<br>$(A \vee B \vee D)(\neg A \vee \neg B \vee \neg D)$<br>$(A \vee C \vee D)(\neg A \vee \neg C \vee \neg D)$<br>$(B \vee C \vee D)(\neg B \vee \neg C \vee \neg D)$ | $\vdash$ | $\neg D$ imply $A \wedge B \wedge C$ |
| (6) $A \oplus B \oplus C \oplus D = 1$<br>$(A \vee B \vee E)$<br>$(A \vee C \vee E)$<br>$(B \vee C \vee E)$ | $\vdash$ | $\neg D \wedge \neg E$ imply $A \wedge B \wedge C$ |
| (7) $A \oplus B \oplus C \oplus D = 1$<br>$\neg A \vee \neg B \vee \neg E$<br>$\neg A \vee \neg C \vee \neg E$<br>$\neg B \vee \neg C \vee \neg E$ | $\vdash$ | $D \wedge E$ imply $\neg A \wedge \neg B \wedge \neg C$ |

The logical consequences here are not a CNF. In order to hand conveniently in them to a general SAT solver, we will encode the logical consequences into a CNF. The logical premise conditions of the above reasoning rules do not necessarily appear in the original SAT problem, but do in some subproblems. During solving, we split the original SAT problem into subproblems, and then extract the above logical premise conditions from the subproblems.

## 3  Algorithm and Implementation

The solver developed here is similar to MoRsat and CircleSAT [6, 21], and is also a hybrid SAT solver, but is built on the top of MXC [12], not Rsat [8, 13]. Although MXC, CryptoMiniSat [20], Rsat and PrecoSAT [9] are a Conflict Driven Clauses Learning (CDCL) solver, based on our empirical observation, MXC is the fastest among these solvers on the hwb problem. In fact, this can be seen from the SAT 2009 competition [11]. Therefore, we decided to use MXC to solve subproblems. The basic idea of our algorithm is to split the original problem into subproblems by a special heuristic, simplify the subproblems, extract structured information from the simplified subproblems, and solve the simplified subproblems with structured information by the MXC solver. The following is the pseudo-code of the algorithm.

**Algorithm 1** SATsolver ($F$, $level$)
  **if** $F$ = empty **then return** satisfiable
  **if** empty clause in $F$ **then return** unsatisfiable
  **if** $level$ = 1 **then**  $P$ := DecisionVar1($F$)
  **if** $level$ = $|P|+4$  **then**
      $F$ := simplify($F$)
      $G$ := {$\beta$ | $\alpha \vdash \beta$ and $\alpha$ in $F$} by rules (1)-(7)
      **return** CDCL_solver($F \cup$ CNF($G$) )
  **endif**
  **if** $level$ < $|P|$ **then** $var$ := unfixed variable in $P$
  **else**  **if** $level$=$|P|$ **then**  $Q$ = DecisionVar2($F$) **endif**
      $var$ := unfixed variable $x$ in $Q$ with greatest $H(x)$
  **if** SATsolver( $F(var$=1), $level$+1) = satisfiable **then return** satisfiable
  **return**  SATsolver($F(var$=0), $level$+1)

**Algorithm 2**  DecisionVar1($F$)
  $P$ := empty
  **for** each $(A \oplus B \oplus C = 1)$ $(A \vee D)(\neg B \vee D)$ in $F$ **do**
      $P := P \cup \{C\}$
      add $\neg C \vee D$ to $F$
  **return** $P$

**Algorithm 3**  DecisionVar2($F$)
  $P$ := empty
  **for** each $(A \oplus B \oplus C = 1)$ $(A \vee D)(\neg B \vee D)$ in $F$ **do**
      $P := P \cup \{A, B, D\}$
  **for** each $x$ in $P$ **do**
      $H(x)$ := occCNF($x$) * occCNF($\neg x$)
  **return** $P$

The algorithm begins with a very simple lookahead process. The heuristic used in the lookahead process is to choose a variable as a decision variable from clauses with the premise condition of rule (1). Except rule (1), the other 6 rules given above do not occur in the original problems tested. Therefore, only rule (1) is used when deciding which variables to branch on. Notice, if rule (1) does not occur, we switch the choice of decision variables to that of the March solver [14, 15], and set $|P|+4$ to 10.  In Algorithm 1, *level* denotes a decision level, and is initialized to be 1. When the decision level is greater than the number of the premise conditions of rule (1), our heuristic choose a variable $x$ in the premise conditions of rule (1) with the graetest $H(x)$. In the March solver [14, 15], $H(x)$ is defined as the product of two ACE (Approximation of the Combined lookahead Evaluation) functions [16]. But $H(x)$ here is simply defined as the product of the number of $x$'s occurrence and $\neg x$'s occurrence in CNF clauses. In Algorithm 3, occCNF($x$) denotes the number of $x$'s occurrences in the CNF formula excluding XOR constraints. When the decision level is low, we must compute $F(x=1)$ or/and $F(x=0)$, where $F(x=1)$ denotes the resulting formula after assigning to true and performing iterative unit propagation, and $F(x=0)$ is similar. On entering the ($|P|+4$)-th decision level, the algorithm invoke a CDCL (Conflict Driven Clauses Learning) solver to solve the subproblem $F \cup$ CNF($G$), where $F$ is a simplified CNF formula, and CNF($G$) is an CNF encoding of the propositional logic consequence $G$. The level of calling a CDCL is set to $|P|+4$. This is based on our experimental observation. Because the general SAT solver handles only a CNF formula, we must encode the propositional logic consequence $G$ obtained from $F$ by rules (1)-(7) into a CNF formula. This encoding work is simple [17, 18]. So we here omit its implementation details. The simplification of $F$ is simpler than SATeLite-like preprocessing. In simplifying $F$, we support only the detection of equivalent literal and the backward subsumption. In the final implementation, CDCL solver is replaced with the MXC solver.

## 4 Empirical evaluation

Table 1. Runtime comparison of three solvers (in seconds). (">10000" shows that the instance cannot be solved in 10000 seconds.)

| Instance | SAT | LogicSAT | MXC 0.99 | Precosat |
|---|---|---|---|---|
| hwb-n24-01 | no | 40.83 | 125.49 | 291.25 |
| hwb-n24-02 | no | 48.04 | 105.44 | 256.72 |
| hwb-n24-03 | no | 36.86 | 96.53 | 166.40 |
| hwb-n26-01 | no | 135.99 | 334.19 | 1292.38 |
| hwb-n26-02 | no | 160.77 | 576.21 | 2059.89 |
| hwb-n26-03 | no | 175.91 | 541.06 | 4142.93 |
| hwb-n28-01 | no | 408.10 | 1047.43 | 2761.41 |
| hwb-n28-02 | no | 815.37 | 1294.80 | >10000 |
| hwb-n28-03 | no | 714.38 | 1330.32 | >10000 |
| hwb-n30-01 | no | 1467.11 | 4966.34 | >10000 |
| hwb-n30-02 | no | 1552.03 | 3601.23 | >10000 |
| hwb-n30-03 | no | 2635.42 | 8497.09 | >10000 |
| hwb-n32-01 | no | 2643.94 | >10000 | >10000 |
| hwb-n32-02 | no | 7305.17 | >10000 | >10000 |
| hwb-n32-03 | no | 4469.18 | >10000 | >10000 |
| lisa21_99_a | yes | 28.75 | 15.13 | 52.43 |
| lisa19_99_a | yes | 28.08 | 43.12 | 2.04 |
| pb-sat-40-4-01 | yes | 280.08 | 1027.08 | 469.57 |
| pb-sat-40-4-02 | yes | 400.01 | 667.41 | 227.08 |
| pb-sat-40-4-03 | yes | 361.98 | 407.28 | 219.14 |
| pb-sat-40-4-04 | yes | 387.62 | 702.71 | 107.48 |
| pb-unsat-40-4-01 | no | 527.60 | 948.32 | 902.34 |
| pb-unsat-40-4-02 | no | 505.34 | 1140.43 | 702.02 |
| pb-unsat-40-4-03 | no | 492.16 | 553.95 | 630.41 |
| pb-unsat-40-4-04 | no | 473.28 | 693.06 | 642.82 |

Based on rules (1) –(7) and algorithms given above, we built a new solver on the top of MXC. This new solver is called LogicSAT. This solver can also be built on the top of other CDCL solvers, say Precosat. However, we found that the solvers built on the top of other solvers were not fast. For example, solving hwb-n32-02 by the solver building on the top of Precosat spent 11205 seconds. We conducted all the experiments on a 2.40GHz machine with Intel Core 2 Quad Q6600 CPU and 2GB RAM, running Mint Linux. As our competitors, we have chosen two latest solvers: Precosat and MXC 0.99 [12]. Precosat won a Gold Medal for the application category at SAT 2009. MXC is not any Gold Medal winner in SAT 2009, but is the fastest on the hwb family, which were selected as an important benchmark, and arises from equivalence checking problem of two different circuits computing the hidden weighted bit function by Stanion [10]. The hwb problems were still used in SAT 2009. So far, no solver can solve the hwb32 problems. In addition to the hwb problems, as our benchmarks, we selected the lisa21-99-a and lisa19_99_a contributed by Aloul [19], and the pb-sat-40 and pb-unsat-40 family contributed by Pyhala Braun to SAT 2002 [19], which is another encoding of factoring problems. The reason why other instances were not selected as our benchmarks is that they cannot apply rules (1)-(7). In our experiments, each solver is given 10000 seconds per problem.

Table 1 reports the runtimes of three solvers on different instances. Compared with MXC, our improvement is significant. Except for lisa21_99_a, our solver, LogicSAT, was consistently faster than MXC. As shown in Table 1, both Precosat and MXC cannot solve the hwb-n32 problems. In fact, to our best knowledge, up to now, no solver can solve them in 10000 seconds. However, our solver solved each of the hwb-n32 problems in 10000 seconds. On the hwb problems, Precosat is not competitive. Even for some of the hwb-n28 problem, Precosat cannot also solve them. Generally speaking, Precosat is worse on unsatisfiable benchmarks, but better on satisfiable benchmarks. For example, for lisa19_99_a and pb-sat, Precosat was faster than our solver. This is because the decision variable selection strategies of the two solvers are different. Not all subproblems can be simplified by applying the logic rules. However, about 90% of subproblems were simplified by applying the logic rules. In general, the typical level at which the underlying CDCL solver is called is from 8 to 12.

## 5 Conclusions

This paper presented a novel approach for SAT problems and solved successfully the hwb-n32 problem, which remains unsolvable so far. The idea of our approach is to split the original problem into some subproblems, exploit logical premise, obtain logical consequences according to logic rules in solving each subproblem, and add the obtained logical consequences to the subproblem. The subproblem with logical consequence clauses can be solved more efficiently than the original subproblem. The experimental results showed that this approach is effective. Because the logic rules used in this paper are few, they can only be applied to a few of SAT problems. To generalize this approach, we must extend the logic rules listed here. How to extend them is still a challenge subject. This is not impossible, since we believe that in addition to the SAT problems listed here, other SAT problems should also contain other logic rules. This depends on real-world SAT problems. As a future research project, we will study how to exploit new logic rules from other SAT problems.